\documentclass[a4paper,twoside,twocolumn,english]{revtex4}
\usepackage[T1]{fontenc}
\usepackage{color}
\usepackage{graphicx}
\usepackage{amssymb}

\makeatletter

\newcommand{\noun}[1]{\textsc{#1}}

\providecommand{\tabularnewline}{\\}

\usepackage{babel}
\makeatother
\begin{document}

\title{Differential Structures -- Geometrization of Quantum Mechanics}

\author{Torsten Asselmeyer-Maluga and Helge Rosé}

\address{FhG FIRST, Kekuléstra{\ss}e 7, 12489 Berlin, Germany\\
torsten@first.fhg.de}

\begin{abstract}
The usual quantization of a classical space-time field does not touch
the non-geometrical character of quantum mechanics. We believe that
the deep problems of unification of general relativity and quantum
mechanics are rooted in this poor understanding of the geometrical
character of quantum mechanics. In \noun{Einstein}'s theory gravitation
is expressed by geometry of space-time, and the solutions of the field
equation are invariant w.r.t. a certain equivalence class of reference
frames. This class can be characterized by the \emph{differential
structure} of space-time. We will show that \emph{matter} is the transition
between reference frames that belong to different differential structures,
that the set of transitions of the differential structure is given
by a \textsc{Temperley-Lieb} algebra which is extensible to a $C^{*}$-algebra
comprising the field operator algebra of quantum mechanics, and that
the state space of quantum mechanics is the linear space of the differential
structures. Furthermore we are able to explain the appearance of the
\emph{complex numbers} in quantum theory. The strong relation to \emph{Loop
Quantum Gravity} is discussed in conclusion.
\end{abstract}

\pacs{04.60.Pp,02.40.Ma,03.65.Ta,04.20.Gz,02.40.Xx}

\keywords{quantum mechanics, differential structure, field operator algebra,
Temperley-Lieb algebra, surgery, Loop quantum gravity, geometrical
structure}

\maketitle

\section{Introduction}

\noun{Einstein}'s theory of general relativity is one of the most
unifying steps in conceivability of nature. In his theory only \emph{two}
entities of being remain: the space-time and matter. And furthermore
these entities are not independent from each other -- they are deeply
connected by Einstein's field equation: The energy of the matter causes
the change of the gravitational field -- a field which is a pure geometrical
property of space-time. In fact, it is matter that shapes the world
-- but matter is delineated by quantum mechanics, which it not captured
by Einstein's theory. Quantum theory pretty much understands the properties
of matter provided there is no genuine concept of space-time requested.
In quantum mechanics space and time are only parameters, the fields
depend on this parametrization, but there is no observable which is
measuring a space-time point.

We believe that the deep problems of unification of general relativity
and quantum mechanics are founded in this poor understanding of the
geometrical character of quantum mechanics. The usual way of quantization
of a classical space-time field does not touch this non-geometrical
character of quantum mechanics. The question will not be: Can we quantize
a geometrical field? The question is: What is the geometry of quantum
mechanics? In other words, the problem is not caused by difficulties
of the quantization -- the problem comes from two contradictory concepts
of explanation. If we are able to explain quantum mechanics by a genuine
geometrical concept, there should be a straight way to the most unifying
theory of physics remaining only \emph{one} entity: \emph{}the \emph{smooth
4-dimensional manifold} of space-time.

Two basic classes of properties classify the structure of a manifold:
topology and geometry. In \noun{Einstein}'s theory gravitation is
expressed by the metrical field, i.e. by a geometrical property. A
short-cut thinking could seduce us to identify matter with the topological
properties of the manifold. Indeed, this would be a nice picture:
The two fundamental physical entities would be described by the two
basic mathematical structures of one manifold. But there is a deep
problem: Topology and geometry are only loosely connected. If the
fundamental theory were based on them, we would not get one, we would
get two dichotomic theories separated forever losing any chance of
unification.

What we need is a \emph{third} structure that lies between topology
and geometry and that has strong connections to both of them. What
we really need is the missing link between both, a structure which
is able to join geometry and topology in single concept. And in fact,
such a third fundamental structure exists: The third class of properties
characterizing a manifold is the \emph{differential structure}.

We will see that differential structures are able to express both,
geometrical and topological properties. But these two characteristics
will be properties of different dimensions: The geometrical character
represents the change of \emph{}connection of a \emph{4-dimensional}
manifold -- it will be the source term in \noun{Einstein}'s equation.
The topological character describes \emph{3-dimensional} singularities,
which can be identified with the particles. Our short-cut thinking
was not completely wrong: Matter \emph{is} characterized by topology,
but it is not the one of a \emph{four} manifold, it is the topology
of \emph{the 3-dimensional} particles of matter. The differential
structure describes matter by two equivalent pictures: a geometrical
one -- as a \emph{change} in the \emph{4-dimensional connection},
and a topological one -- as \emph{3-dimensional singularities}.

These two pictures of the differential structure naturally express
the dualism of the complementary aspects of matter in quantum mechanics:
the \emph{non-local, wave-like} behavior of the 4-dimensional field
of connection, and the \emph{local, particle-like} character of 3-dimensional
singularities. In this way, the differential structure not only unifies
topology and geometry --~read: matter and gravitation~--, it furthermore
\emph{explains} the dualism of quantum mechanics by the \emph{mathematical}
fact that its non-local 4-dimensional field of connection is equivalently
described by local 3-dimensional singularities.

Including all this: If the concept of differential structures is appropriate,
it should be able to generate the structures of quantum mechanics
naturally, i.e. the Hilbert space and the algebra of field operators.
In this article we will show that the differential structure is not
merely able to lead to the Hilbert space and the field operator algebra
-- these basic structures of quantum mechanics are the inherent expressions
of the very meaning of the differential structure. And furthermore,
the differential structure is exceptional only for one case: only
a \emph{4-dimensional} manifold has an \emph{infinite} number of differential
structures.

Up to now, there is no complete mathematical theory of differential
structures, so let us start with a physical point of view of differential
structures. A manifold is described by charts $h_{i}$: homeomorphic
maps from subsets of the manifold $M$ into the linear space $\mathbb{R}^{n}$\[
h_{i}\colon M\supset W_{i}\rightarrow U_{i}\subset\mathbb{R}^{n}.\]

These charts describe the local properties of the manifold captured
by linear spaces. But the really interesting property is the structure
between these charts. Assume two charts $h_{i}\colon W_{i}\rightarrow U_{i}$
and $h_{j}\colon W_{j}\rightarrow U_{j}$. The overlapping origin
$W_{ij}=W_{i}\cap W_{j}$ will be mapped into two (usually different)
images $U_{ij}=h_{i}\left(W_{ij}\right)$ and $U_{ji}=h_{j}\left(W_{ij}\right)$.
A \emph{coordinate transformation} between two charts is a map between
subsets of linear spaces: \[
h_{ij}\colon U_{ij}\rightarrow U_{ji},\,\, h_{ij}(x)=h_{j}\left(h_{i}^{-1}\left(x\right)\right).\]

Two charts $h_{i},\, h_{j}$ are \emph{compatible} if $U_{ij},\, U_{ji}$
are open (possibly empty), and the coordinate transformations $h_{ij},\, h_{ji}$
(with $W_{i}\cap W_{j}\neq\emptyset$) are \emph{diffeomorphisms}.
A family of pairwise compatible charts that covers the whole manifold
is an \emph{atlas,} and two atlases are \emph{equivalent} if their
union is an atlas again. The \emph{equivalence classes} of atlases
are the \emph{differential structures} of the manifold.

In physics, charts are the \emph{reference frames}. A coordinate transformation
from one frame to a compatible one is a diffeomorphism for which \noun{Einstein}'s
equation is invariant. Thus all solutions of \noun{Einstein}'s equation
in reference frames of the same differential structure are equivalent.
But a transition between reference frames of different differential
structures leads to non-equivalent solutions of \noun{Einstein}'s
equation, i.e. acts like a source term of the equation. These sources
represent matter. That means in the geometrical picture: \emph{Matter
is the transition between reference frames that belong to different
differential structures of space-time}.

The first discussion of differential structures appeared in a series
of papers \cite{BraRan:93,Bra:94a,Bra:94b} written by \noun{Brans.}
A further relation to particle physics was discussed in \cite{Sla:96,Sla:96b,Sla:96c}.
By using non-standard analysis and logic, \noun{Krol} \cite{Krol:04a,Krol:04b}
conjectured a strong connection between quantum mechanics and differential
structures. Furthermore, in \cite{Bra:94a} \noun{Brans} conjectured
about sources of gravity given by differential structures (Brans conjecture)
which was proven for some compact manifolds by one of authors \cite{Ass:96}
and for some non-compact spaces in \cite{Sla:99}. In the following
we are mainly interested in the transition of the differential structure.
A detailed investigation of the transition of a differential structure
was given in \cite{Ass:96}. The basic idea is simple: The Levi-Civita
connection of a manifold $M$ depends deeply on the differential structure,
and a transition of the differential structure will cause a change
of the connection. Vice versa, a change of the Levi-Civita connection
which is not induced by a diffeomorphism will result in a transition
of the tangent bundle. The local trivialization of that bundle is
the differential structure of the manifold $M$. Let $\omega$ be
the Levi-Civita connection corresponding to a given differential structure.
A diffeomorphism $g\in{\it Diff}(M)$ will change this connection
$\omega$ to\[
\omega'=g^{-1}\omega\, g+g^{-1}dg\,.\]
 The corresponding curvature $R=d\omega+\omega\wedge\omega$ is transformed
to\[
R'=g^{-1}R\, g\]

\[
d\omega'+\omega'\wedge\omega'=g^{-1}(d\omega+\omega\wedge\omega)\, g\,.\]
Because of the relation\[
d(g^{-1}dg)+g^{-1}dg\wedge g^{-1}dg=0\]
the diffeomorphism $g$ does \emph{not} produce an additional curvature.
In the next section we will show that a map $f\colon M\rightarrow M$
which is \emph{smooth}, but has \emph{singularities} along a (closed)
\emph{3-dimensional} submanifold $\Sigma$ will change the differential
structure of $M$. We denote by $f_{*}$ the group element induced
by the differential \textcolor{black}{$df\colon TM\to TM$. If the
map $f$ admits singularities then the differential $df$ and the
group element $f_{*}$ are not well-defined. But it is possible to
define the connection and the curvature corresponding to the singular
map $f$. We omit this technical detail here and refer to the appendix.
Then} the corresponding change of the connection yields\[
\omega'=f_{*}^{-1}\omega\, f_{*}+f_{*}^{-1}df_{*}\]
 with the transformed curvature\[
R'=d\omega'+\omega'\wedge\omega'=f_{*}^{-1}R\, f_{*}+d(f_{*}^{-1}df_{*})+f_{*}^{-1}df_{*}\wedge f_{*}^{-1}df_{*}\,.\]
 The additional curvature $d(f_{*}^{-1}df_{*})+f_{*}^{-1}df_{*}\wedge f_{*}^{-1}df_{*}$
leads to a source term in \noun{Einstein}'s equation \cite{Ass:96}:
A transition of the differential structure will change the source-free
field equation %
\footnote{In the following we will present the field equation coordinate independent
for arbitrary vector fields $X,Y$.%
}\[
Ric(X,Y)-\frac{1}{2}\, g(X,Y)R=0\]
 to\[
Ric(X,Y)-\frac{1}{2}\, g(X,Y)R=\frac{w}{w+1}\, R\left(f_{*}^{-1}df_{*}\right)(X,Y)\]
 where $R((f_{*}^{-1}df_{*}))$ is the Ricci curvature part of the
singular curvature induced by the singular map $f$ and $w$ is the
winding number of $f$.\\
The energy density $T_{00}$ of the energy-momentum tensor $T$ of
a particle with mass $m$ and a volume $vol(\Sigma)$ of the manifold
$\Sigma$ is given by \[
T_{00}=mc^{2}\,\delta_{\Sigma}\]
 where $\delta_{\Sigma}$ is Diracs delta function with support on
$\Sigma$, i.e\[
{\displaystyle \int\limits _{\Sigma}}\delta_{\Sigma}=1\]
By the normal form of the field equation\[
Ric(X,Y)-\frac{1}{2}\, g(X,Y)R=\frac{8\pi G}{c^{4}}\, T(X,Y)\,.\]
 we get a relation between the energy and the geometry. The total
energy of the particle is given by the integral\[
\int\limits _{\Sigma}T_{00}\: dvol(\Sigma)=mc^{2}\]
 and by the field equation we get for the \emph{mass} of a particle

\[
m=\frac{c^{2}}{8\pi G}\,\frac{w}{w+1}\,\int\limits _{\Sigma}\: R\left(f_{*}^{-1}df_{*}\right)_{00}\: dvol(\Sigma)\]
 where $R(f_{*}^{-1}df_{*})_{00}$is the $00$-component of the Ricci
tensor generated by the singular form $f_{*}^{-1}df_{*}$. In section
\ref{sec:4} we will calculate the singular curvature, which corresponds
to a closed 2-manifold with area $A$ via Poincar\'e duality. It
is interesting to note that the area of the 2-manifold is fixed %
\footnote{In 3-manifold topology, such a 2-manifold is called an incompressible
surface.%
} with respect to the volume of $\Sigma$. Then the length $\lambda(f)=vol(\Sigma)/A$
depends only on the map $f$. After this simple argumentation, we
can express the \emph{mass} of the particle by\[
m(f)=\frac{c^{2}}{8\pi G}\,\frac{w}{w+1}\:\lambda(f)\,,\,\,\lambda(f)=\frac{vol(\Sigma)}{A}.\]
In fact we have obtained by a simple calculation: A \emph{massive,
extended particle} as the very consequence of the \emph{transition}
of the differential structure\emph{.}\\
\\
Sure, this is a simplification. Especially there is no explanation
yet for the quantum nature of the particles. In Section \ref{sec:3}
we will give a complete explanation by constructing the field operator
algebra of quantum mechanics directly from the set of transitions
of differential structure. We will explicitly show that differential
structures on 4-manifolds are strongly connected to (immersed) surfaces.
Such surfaces can be described by a principal $U(1)$ bundle over
the four-manifold. That is the main reason for the appearance of complex
numbers in quantum mechanics. The intersections between these surfaces
are described by the defining relations of a Temperley-Lieb algebra.
In Section \ref{sec:4} we will construct concrete examples of different
differential structures. We also will illuminate the necessity of
a complex Hilbert space in quantum mechanics by an explicit construction
of the amplitudes and phases of the quantum states. In the last Section
\ref{sec:6} we will discuss a relation between our approach and the
loop quantum gravity.

\section{Differential structures on four-manifolds}

A \emph{differential structure} of the manifold $M$ is an \emph{equivalence
class} of the atlases of the manifold $M$. We call two atlases ${\mathcal{A,A'}}$
\emph{equivalent} iff there are \emph{diffeomorphisms} between the
transition functions. As an important fact we will note that there
is only \emph{one} differential structure of any manifold of dimension
smaller than \emph{four}. For all manifolds larger than four dimensions
there is only a \emph{finite} number of possible differential structures
$\mbox{{\it Diff}}{}_{\dim M}$. The following table lists the numbers
of differential structures up to dimension $11$.\\

\begin{tabular}{|c|c|c|c|c|c|c|c|c|c|c|c|}
\hline 
n&
1&
2&
3&
4&
5&
6&
7&
8&
9&
10&
11\tabularnewline
\hline
\hline 
$\#{\it Diff}_{n}$&
1&
1&
1&
$\infty$&
1&
1&
28&
2&
8&
6&
992\tabularnewline
\hline
\end{tabular}\\
\\
In dimension \emph{four} there is a countable number of differential
structures on most compact four-manifolds and an uncountable number
for most non-compact four-manifolds.

To get a first view at the terra incognita the space of 4-dimensional
differential structures builds we need a better definition of a differential
structure. A first clue is provided the fact that a \emph{diffeomorphism}
$M\rightarrow N$ between two manifolds $M,\, N$ induces an i\emph{somorphism}
between the corresponding tangent bundles $TM,\,\, TN$. A transition
between two only topological (i.e. homeomorphic) equivalent manifolds,
i.e. with different differential structures, causes inequivalent tangent
bundles. The next hint says that the differential structure is also
encoded into the structure of the algebra $C^{\infty}(M)$ of smooth
functions $M\rightarrow{\mathbb{C}}$. This fact is connected with
deformations of the algebra $C^{\infty}(M)$ providing an algebra
of the different differential structure. Such deformations are studied
in non-commutative geometry. We will use a combination of both approaches:
We will construct for the transition of the differential structure
the non-commutative algebra of deformations of $C^{\infty}(M)$ and
the transition of the tangent bundle described by the change of the
connection. And we will formalize this approach by considering two
homeomorphic four-manifolds $M,\, N$ with different differential
structures.

Unfortunately it is too difficult to use the homeomorphism $M\rightarrow N$
to construct the transition of the tangent bundle $TM\rightarrow TN$.
The tricky way is to use a smooth map $f\colon M\rightarrow N$ which
is surjective but not injective %
\footnote{Otherwise the smooth map is bijective and thus a diffeomorphism.%
}. Such a map is \emph{singular}, i.e. the linear map of the differential
\textcolor{black}{\[
df_{x}\colon T_{x}M\rightarrow T_{f(x)}N\]
 has not the maximal rank and we define the} \textcolor{black}{\emph{singular
set}} \textcolor{black}{$\Sigma$ to be \[
\Sigma={x\in M\:|\: rank(df_{x})<4}\:.\]
 It is obvious that the transition of the differential structure is
strongly related to the structure of the singular set $\Sigma$. The
explanation of this fact in the next section will show the geometrical
nature of the quantum incarnated in the transition of differential
structure. }

\section{Construction of the field operator algebra }

\label{sec:3}The fundamental hypothesis of this paper is the statement
that the structures of quantum mechanics are naturally induced by
the differential structure of the space-time-manifold. In this section
we will justify this by proving the

\begin{description}
\item [Theorem\label{Theorem1}:]\emph{Let} $\mathcal{T}$ \emph{be the
differential} $*$\emph{-algebra of singular 1-forms defined below
forming the set of transitions of the differential structure. Furthermore,
let} $D_{\varphi}$ \emph{be the covariant derivative associated to
the singular form} $\varphi$ \emph{and the universal derivative}
$L$ \emph{of} $\mathcal{T}$ \emph{defined by} $\varphi L\psi:=D_{\varphi}\psi$.
\emph{Then the algebra} $\mathcal{T}$ \emph{has the structure of
a Temperley-Lieb algebra. The algebra of transitions} $\mathcal{T}$
\emph{is extensible to a} $C^{*}$\emph{-algebra comprising the algebra
of field operators of particles. The complex Hilbert space is induced
by} $\mathcal{T}$ \emph{via the GNS construction.}
\end{description}
In the introduction we have shown that there is a close relation between
the transition of the differential structure and a singular connection
with 3-dimensional support. Such connections are expressed by singular
1-forms with 3-dimensional supports. At a first glance, one might
be tempted to assume that the vector space spanned by these 1-forms
is capable of describing all transitions of the differential structure.
However, as it will turn out, it is the \emph{support} of the 1-forms
which is important. Thus we are looking for a structure that captures
the transitions of the support of the singular 1-forms. Supports,
in their capacity of being sets, suggest two operations, viz. union
and intersection. Thus we would expect that the transitions of the
differential structure are described by an algebraic structure with
two operations, i.e., by an algebra. This is what we are going to
prove in the sequel.

We will confine ourself to the case where the tangent bundle is changed
only by a variation of the differential structure. But a transition
of the tangent bundle is caused by the change of its connection. Thus,
we have to discuss the effect of a singular smooth map that changes
the connection.

Given a singular smooth map $f\colon M\to N$ representing a variation
of the four-manifold $M$ and causes a transition of the differential
structure. Let $D_{M}$ and $D_{N}$ be the covariant derivatives
of the tangent bundles $TM$ and $TN$. Any covariant derivative can
be decomposed by $D=d+\omega$ with the connection 1-form $\omega$.
We get the transformation \[
D_{M}=f_{*}^{-1}D_{N}\, f_{*}\]
of $D_{M}$ to $D_{N}$ with the group element $f_{*}$ induced by
the differential $df\colon TM\to TN$ and the connection 1-form transforms
by\[
\omega_{M}=f_{*}^{-1}\omega_{N}\, f_{*}+f_{*}^{-1}df_{*}\,.\]
 The inhomogeneous contribution of the change of the connection is
given by the \emph{singular 1-form} \[
\varphi=f_{*}^{-1}df_{*}\]
 which has a support at the singular set $\Sigma$ of $f$. Before
we proceed with the definition of a singular 1-form, we have to discuss
the action of a diffeomorphism. Let $g\colon M\to M$ be a diffeomorphism
that induces a map $dg\colon TM\to TM$ which is also a diffeomorphism.
The application of $g$ changes the derivative $D$ to $g_{*}^{-1}D\, g_{*}$
and keeps the trace $Tr_{M}(D)$ of operator $D$ over $M$ invariant:
$Tr_{M}(g_{*}^{-1}D\, g_{*})=Tr_{M}(D)$ for all diffeomorphisms $g$.
The last relation is only true for diffeomorphisms and not for singular
maps like $f$. Thus instead of using the operator~$D$, which is
not diffeomorphism-invariant, we consider the diffeomorphism-invariant
trace $Tr_{M}(D)$. 

There are two methods to define the singular form $\varphi$:

\begin{enumerate}
\item Given the two connection 1-forms $\omega_{M},\:\omega_{N}$ defined
above, then\[
\varphi=\omega_{M}-f_{*}^{-1}\omega_{N}\: f_{*}\]

\item or consider the exterior differential $d$ and define\[
\varphi=D-d\qquad D=f_{*}^{-1}(d)f_{*}=d+f_{*}^{-1}df_{*}\]

\end{enumerate}
The first method was used in \cite{Ass:96} to calculate the change
of free \noun{Einstein} equation caused by the transition of differential
structure. In this paper we study the second method. The question
is: What is the difference between $d$ and $D$? By the theory of
differential forms we know that $d$ is the exterior derivative with
$d^{2}=0$. In contrast to this, $D$ is a more general derivative:
$D$ fulfills the Leibniz rule $D(\varphi\psi)=(D\varphi)\psi+\varphi(D\psi)$
\emph{and} $D^{3}=0$ in place of $d^{2}=0$. The square $D^{2}=d\varphi+\varphi\wedge\varphi$
-- in general \emph{not} zero -- is the \emph{curvature} of the singular
connection change $\varphi$. The relation $D^{3}=0$ is the Bianchi
identity. If we express the curvature $D\varphi$ as the covariant
derivative of $\varphi$, then the Bianchi identity is given by $D^{2}\varphi=0$.
That fact is used below to construct an exterior derivative from the
covariant derivative. Thus, a generating element of the connection
change is given by the 1-form $\varphi$ with $supp(\varphi)=\Sigma_{\varphi}$
which is the singular set $\Sigma_{\varphi}$ of the smooth map $f\colon M\rightarrow N$.

In the following we will show that the set $\mathcal{S}$ of singular
1-forms $\varphi$ has an algebra structure. Its two defining operations
-- sum and product -- are related to the union and intersection of
the supports. Thus we can define the \emph{sum} by\[
\varphi+\psi=\chi\,:\,\varphi,\psi,\chi\in\mathcal{S},\,\, supp(\chi)=supp(\varphi)\cup supp(\psi)\,.\]
 The construction of the product $\varphi\cdot\psi$ is a bit more
complicated. Given two forms $\varphi,\:\psi$ with their singular
supports $\Sigma_{\varphi},\:\Sigma_{\psi}$. Any 1-form is naturally
associated with a curve lying in its singular support: This \emph{associated
curve} $C_{\varphi}$ is defined by the Poincar\'e dual of the 1-form
$\varphi$ \textcolor{black}{(see \cite{Bre:93}), i.e. via the integral
of the 1-form\[
\int\limits _{C_{\phi}}\phi=1\:.\]
 The curves of $\varphi,\:\psi$ are generators of the fundamental
groups $\pi_{1}(\Sigma_{\varphi}),\:\pi_{1}(\Sigma_{\psi})$.} In
general, for a map $f:\: M\to N$ with singular set $\Sigma$ there
are more than one generator of the fundamental group $\pi_{1}(\Sigma)$.
Let $C_{\varphi}$ and $C_{\psi}$ be two generators of $\pi_{1}(\Sigma)$.
Furthermore, let $S_{\varphi}$ be the surface with boundary $\partial S_{\varphi}=C_{\varphi}$
also known as Seifert surface. Because of the knotting of the curve
$C_{\varphi}$, the Seifert surface can be very complicated (see \cite{Rol:76}).
The two curves $C_{\varphi},\: C_{\psi}$ are \emph{linked} if their
corresponding surfaces intersect transversally. We denote linked curves
by $C_{\varphi}\between C_{\psi}$ and the transversally intersection
by $S_{\varphi}\pitchfork S_{\psi}$. By this, we are able to define
the \emph{product}\[
\varphi\cdot\psi=\chi\,:\,\varphi,\psi,\chi\in\mathcal{S},\,\, supp(\chi)=supp(\varphi)\cap supp(\psi)\]
 where the product curve $C_{\chi}$ is given by\[
C_{\chi}=\left\{ \begin{array}{lcl}
C_{\varphi}\between C_{\psi} & : & S_{\varphi}\pitchfork S_{\psi}\\
C_{\varphi}\sqcup C_{\psi} & : & S_{\varphi}\cap S_{\psi}=\emptyset\end{array}\right.\]
The first case $C_{\varphi\cdot\psi}=C_{\varphi}\between C_{\psi}$
represents a \emph{non-commutative} product $\varphi\cdot\psi\not=\psi\cdot\varphi$
because the link $C_{\psi\cdot\varphi}=C_{\psi}\between C_{\varphi}$
is different from $C_{\varphi\cdot\psi}$. In knot theory one calls
$C_{\psi\cdot\varphi}$ the mirror link of $C_{\varphi\cdot\psi}$.
In the second case one gets from $C_{\varphi\cdot\psi}=C_{\varphi}\sqcup C_{\psi}$
by using $C_{\varphi}\sqcup C_{\psi}=C_{\psi}\sqcup C_{\varphi}$
the relation $C_{\varphi\cdot\psi}=C_{\psi\cdot\varphi}$. Thus, in
this case the product is \emph{commutative} $\varphi\cdot\psi=\psi\cdot\varphi$.
The product with a number field $K$ is induced from the corresponding
operation of differential forms. That completes the construction.\\
\textcolor{red}{}\\
\textcolor{black}{The set of singular 1-forms $\mathcal{S}$ endowed
with these two operations forms $\mathcal{T}=(\mathcal{S},+,\cdot)$
-- the} \textcolor{black}{\emph{algebra of transitions of the differential
structure of space-time}}\textcolor{black}{, i.e. of transitions between
non-diffeomorphic reference frames. In the introduction we have motivated
that the transitions $\varphi\in\mathcal{T}$ of the differential
structure represent} \textcolor{black}{\emph{matter}} \textcolor{black}{by
generating additional source terms in the gravitational field equation.
In the following we will show that this algebra $\mathcal{T}$ of
transitions of differential structure of space-time also has a direct
meaning in quantum mechanics: $\mathcal{T}$ is the algebra of field
operators of particles -- the} \textcolor{black}{\emph{Teilchen}}\textcolor{black}{-algebra.}\textcolor{red}{}\\
\textcolor{red}{}\\
For each singular 1-form $\varphi\in\mathcal{T}$ there is a natural
operator $D_{\varphi}\colon\mathcal{T}\to\Omega^{1}(\mathcal{T})$
-- called the \emph{covariant derivative} w.r.t. to $\varphi$ --
mapping an element $\psi$ of $\mathcal{T}$ to an 1-form $D_{\varphi}\psi\in\Omega^{1}(\mathcal{T})$
over $\mathcal{T}$ satisfying the Leibniz rule $D_{\varphi}(\psi\cdot\chi)=(D_{\varphi}\psi)\cdot\chi+\psi\cdot(D_{\varphi}\chi)$.
The expression \[
\Phi=D_{\varphi}\varphi,\,\Phi\in\Omega^{1}(\mathcal{T})\]
 is called the \emph{curvature} of $\varphi$. According to \cite{Bou:89},
every algebra admits an universal derivative $L$ with $L^{2}=0$
which also makes $\mathcal{T}$ to a differential algebra. Furthermore,
we can define formal differential forms $\Omega^{p}(\mathcal{T})$
on $\mathcal{T}$ by using $L$. Then, a $p$-form is generated by
$\varphi_{0}L\varphi_{1}\cdots L\varphi_{p}$ or $\mathbf{1}L\varphi_{1}\cdots L\varphi_{p}$
(see \cite{Con:95}). Every two elements $\varphi,\psi\in\mathcal{T}$
define an element of $\Omega^{1}(\mathcal{T})$ by $D_{\varphi}\psi$.
By the universal property of the derivative $L$, $\Omega^{1}(\mathcal{T})$
is generated by forms like $\varphi L\psi$ or $\mathbf{1}L\varphi$.
Thus, also $D_{\varphi}\psi$ must be given by a form like this and
we may choose \begin{equation}
D_{\varphi}\psi=\varphi L\psi\,.\label{eq:rel-D-L}\end{equation}
In contrast to the usual representation we do not have a wedge product
$\wedge$ and thus we cannot define the covariant derivative like
$D_{\varphi}=L+\varphi\wedge$. But we can choose the derivative $L$
in such a way (using the universality property of $L$) that the relation
$D_{\varphi}\psi=\varphi L\psi$ is fulfilled. Then the curvature
can be written as $\Phi=D_{\varphi}\varphi=\varphi L\varphi$.

\textcolor{black}{Furthermore, we can introduce a} \textcolor{black}{\emph{trace}}
\textcolor{black}{of a singular form $\varphi=f_{*}^{-1}df_{*}$ with
respect to a curve $C$ by the integral\[
Tr(\varphi,C):=\int\limits _{C}\varphi<\infty\:.\]
 The universal derivative $L$ extends the trace over $\mathcal{T}$
to all forms $\Omega^{p}(\mathcal{T})$ by relation (\ref{eq:rel-D-L}).
Using general properties of $L$, one obtains the relation }

\textcolor{black}{\begin{equation}
Tr(\Phi\Psi,C)=(-1)^{pq}Tr(\Psi\Phi,C),\,\,¸\Phi\in\Omega^{p}({\mathcal{T}}),\,\Psi\in\Omega^{q}({\mathcal{T}})\label{graduate-trace}\end{equation}
 for the trace on $\mathcal{T}$ with a suitable curve $C$.} The
construction of one operation is unsettled: the star operation $*$
which makes $\mathcal{T}$ to a $*$-algebra. A $*$-operation has
to fulfill the two relations: $(\varphi^{*})^{*}=\varphi$ and $(\varphi\cdot\psi)^{*}=\psi^{*}\cdot\varphi^{*}$.
Let $\varphi$ be a singular 1-form and $C_{\varphi}$ the corresponding
curve. We define $\varphi^{*}$to be a 1-form with curve $C(\varphi^{*})=\overline{C_{\varphi}}$,
i.e. the curve $C_{\varphi}$ with the opposite orientation. Then
the first relation $(\varphi^{*})^{*}=\varphi$ is obvious. The second
relation $(\varphi\cdot\psi)^{*}=\psi^{*}\cdot\varphi^{*}$ is a standard
fact from knot theory: the change of the orientation of a link transforms
the link to the mirror link. Let $\varphi\cdot\psi$ be a link then
$(\varphi\cdot\psi)^{*}$ is the mirror link. By definition, $\psi^{*}\cdot\varphi^{*}$
is also the mirror link. That completes the construction of the algebra
$\mathcal{T}$.

Finally we obtain:

\begin{enumerate}
\item $\mathcal{T}$ is a differential $*$-algebra $\mathcal{T}=(\mathcal{S},+,\cdot,L)$,
\item \textcolor{black}{the covariant derivative $D_{\varphi}$ is related
to $L$ by $D_{\varphi}\psi=\varphi L\psi$,}
\item \textcolor{black}{the finite trace $Tr(\varphi,C)=\int_{C}\varphi$
fulfills relation (\ref{graduate-trace}).}
\end{enumerate}
\textcolor{black}{In the sequel we will show that the algebra $\mathcal{T}$
is a Tem\-per\-ley-Lieb algebra} over a number field which are the
complex numbers. For the special case, $\mathcal{T}$ will be identical
to the field operator algebra of the fermions. That proves that the
algebra of transitions of differential structures leading to the main
structure of quantum field theory.

A \emph{Temperley-Lieb algebra} is an algebra with unit element $\mathbf{1}$
over a number field $K$ generated by a countable set of generators
$\{ e_{1},e_{2},\ldots\}$ with the defining relations\begin{eqnarray}
e_{i}^{2}=\tau\cdot e_{i}\,, & e_{i}e_{j}=e_{j}e_{i}\,:\,|i-j|>1,\nonumber \\
e_{i}e_{i+1}e_{i}=\tau e_{i}\,, & e_{i+1}e_{i}e_{i+1}=\tau e_{i+1}\,,\, e_{i}^{*}=e_{i}\label{Jones-algebra}\end{eqnarray}
 where $\tau$ is a real number in $(0,1]$. By \cite{Jon:83}, the
Temperley-Lieb algebra has a uniquely defined trace $Tr$ which is
normalized to lie in the interval $[0,1]$. The proof is carried out
by the following steps:

\begin{itemize}
\item The \emph{curvature} of the sum of two elements (\ref{sum-operation})
consists of four terms which have to be interpreted independently.
This can be done by means of the product of $\mathcal{T}$. 
\item Using the set-theoretical formula (\ref{set-formula}), we will get
relations (\ref{rel2}) between some special products of elements.
These products are interpreted as operations between the supports
of the singular forms.
\item The singular form is a change of connection. Every connection has
to fulfill the Bianchi identity. This identity is expressed as $B(\varphi,\varphi,\varphi)=0$
for the trace (\ref{invariant2}). We have to check that all curves
$\varphi(t)$ in $\mathcal{T}$ together with all infinitesimal variations
(\ref{deformation}) of the curve leaving invariant the Bianchi identity
$B(\varphi,\varphi,\varphi)=0$. This results in the fact, that the
\emph{basis elements} $e_{i}$ of the algebra $\mathcal{T}$ are \emph{projection
operators}.
\item It is possible to introduce an \emph{order structure} on the set of
basis elements. If the four-manifold is compact then the basis set
is countable.
\item Finally, it is possible to introduce a star operation and to show
that the number field are the complex numbers.
\end{itemize}
In the first step of the proof we will study the curvature of the
sum of two singular forms $\varphi,\,\psi$, i.e. the expression $D_{\varphi+\psi}(\varphi+\psi)$: 

\begin{eqnarray}
D_{\varphi+\psi}(\varphi+\psi) & = & (\varphi+\psi)L(\varphi+\psi)\label{sum-operation}\\
 & = & (\varphi L\varphi+\psi L\psi)+(\varphi L\psi+\psi L\varphi)\,.\end{eqnarray}
 Each of the four summands of the right-hand side of (\ref{sum-operation})
has to be expressible as a singular form of its own. We know that
$\varphi L\varphi$ and $\psi L\psi$ are the curvatures of $\varphi$
and $\psi$, respectively. Furthermore, the sum of two singular forms
$\varphi,\:\psi$ with non-intersecting supports produces the sum
of curvatures $D_{\varphi+\psi}(\varphi+\psi)=D_{\varphi}\varphi+D_{\psi}\psi$.
But, how can we interpret the expressions $D_{\varphi}\psi=\varphi L\psi,\: D_{\psi}\varphi=\psi L\varphi$?
Let us consider the following construction. Non-intersecting supports
$\Sigma_{\varphi}',\,\Sigma_{\psi}'$ can be constructed from $\Sigma_{\varphi},\:\Sigma_{\psi}$
by\[
\Sigma_{\varphi}'=\Sigma_{\varphi}\setminus(\Sigma_{\varphi}\cap\Sigma_{\psi})\,,\qquad\Sigma_{\psi}'=\Sigma_{\psi}\setminus(\Sigma_{\varphi}\cap\Sigma_{\psi})\,.\]
 and we get the set-theoretical formula:\begin{equation}
\Sigma_{\varphi}\cup\Sigma_{\psi}=(\Sigma_{\varphi}'\sqcup\Sigma_{\psi}')\cup(\Sigma_{\varphi}\cap\Sigma_{\psi})\label{set-formula}\end{equation}
 where $\Sigma_{\varphi}'\sqcup\Sigma_{\psi}'$ is the disjoint union,
i.e. the union $\Sigma_{\varphi}'\cup\Sigma_{\psi}'$ with $\Sigma_{\varphi}'\cap\Sigma_{\psi}'=\emptyset$.
By the arguments above, we identify $\varphi L\varphi+\psi L\psi$
with the curvature of a singular form of support $\Sigma_{\varphi}'\sqcup\Sigma_{\psi}'$.
Thus we have to associate the cross term $\varphi L\psi+\psi L\varphi$
to a singular form of support $\Sigma_{\varphi}\cap\Sigma_{\psi}$.
By construction, that is the product $\cdot$ in the algebra $\mathcal{T}$.
We obtain for the curvature of the product:

\[
D_{\varphi\cdot\psi}(\varphi\cdot\psi)=\varphi\cdot\psi L(\varphi\cdot\psi)=\varphi\cdot\psi(L\varphi)\psi+\varphi\cdot\psi\cdot\varphi(L\psi)\]
 using the Leibniz rule of the derivative $L$. With (\ref{graduate-trace}),
the trace of the curvature of the product is given by\begin{equation}
Tr(\varphi\cdot\psi L(\varphi\cdot\psi))=Tr(\psi\cdot\varphi\cdot\psi\: L\varphi)+Tr(\varphi\cdot\psi\cdot\varphi\: L\psi)\,.\label{eq:cur_prod}\end{equation}
 We obtain the meaning of the terms in (\ref{sum-operation}) by making
the ansatz of the products\begin{eqnarray}
\psi\cdot\varphi\cdot\psi=\tau\psi\,, & \varphi\cdot\psi\cdot\varphi & =\tau\varphi\label{rel2}\end{eqnarray}
 for the intersection $\Sigma_{\varphi}\cap\Sigma_{\psi}\not=\emptyset$.
Inserting (\ref{rel2}) in (\ref{eq:cur_prod}) we get the cross term
$\varphi L\psi+\psi L\varphi$. But this is exactly the interpretation
of the addition (\ref{sum-operation}) by using the set-theoretical
formula (\ref{set-formula}). As noted above, the product $\varphi\cdot\psi$
can be commutative or not. For the commutative case, we obtain a restriction
by relation (\ref{rel2}): $\varphi\cdot\psi^{2}=\tau\psi$. A simple
manipulation shows $(\varphi\cdot\psi)^{2}=\tau\varphi\cdot\psi$.
Thus, the product $\varphi\cdot\psi$ is a projection operator.

With (\ref{eq:rel-D-L}) the Bianchi identity is given by \begin{equation}
D_{\varphi}^{2}\varphi=\varphi L\varphi L\varphi=0\label{eq:Bianchi-identity}\end{equation}
 for any element $\varphi\in\mathcal{T}$. Thus each element of an
1-parameter family $\varphi(t)$ -- a curve in $\mathcal{T}$ -- has
to keep invariant the identity (\ref{eq:Bianchi-identity}) with respect
to any deformation\begin{equation}
\left.\frac{d}{dt}\varphi(t)\right|_{t=0}=[\psi,\varphi(0)]=\psi\varphi(0)-\varphi(0)\psi\label{deformation}\end{equation}
 for some element $\psi\in\mathcal{T}$. This leads to the fact that
the product in $\mathcal{T}$ is not determined completely. An \emph{infinitesimal
deformation} of the \emph{}product to a new one is just defined by
(\ref{deformation}) \cite{Con:95}. This equation has to be fulfilled
for \emph{all} deformations because the Bianchi identity is valid
for all differential structures. To study the consequences \textcolor{black}{of
this demand we define the traces\begin{eqnarray}
A(\varphi_{0},\varphi_{1}) & := & {\mbox{T}r}(\varphi_{0}L\varphi_{1},C)\,,\label{invariant1}\\
B(\varphi_{0},\varphi_{1},\varphi_{2}) & := & {\mbox{T}r}(\varphi_{0}L\varphi_{1}L\varphi_{2},C)\,,\label{invariant2}\end{eqnarray}
 for suitable curves $C$, fulfilling the relations} \begin{eqnarray}
A(\varphi_{0},\varphi_{1}) & = & A(\varphi_{1},\varphi_{0})\label{cyclic1}\\
A(\varphi_{0}\varphi_{1},\varphi_{2}) & - & A(\varphi_{0},\varphi_{1}\varphi_{2})=0\,,\label{symmetric1}\end{eqnarray}
 and\begin{eqnarray}
B(\varphi_{0},\varphi_{1},\varphi_{2}) & = & B(\varphi_{2},\varphi_{0},\varphi_{1})=B(\varphi_{1},\varphi_{2},\varphi_{0})\\
B(\varphi_{0}\cdot\varphi_{1},\varphi_{2},\varphi_{3}) & - & B(\varphi_{0},\varphi_{1}\cdot\varphi_{2},\varphi_{3})+\nonumber \\
B(\varphi_{0},\varphi_{1},\varphi_{2}\cdot\varphi_{3}) & - & B(\varphi_{3}\cdot\varphi_{0},\varphi_{1},\varphi_{2})=0,\label{symmetric2}\end{eqnarray}
where we have used (\ref{graduate-trace}). By means of the trace
$B$ the \emph{Bianchi identity} becomes \begin{equation}
B(\varphi,\varphi,\varphi)=0\,.\label{eq:bianchi}\end{equation}
For the curve $\varphi(t)$ we demand that all infinitesimal deformations
of $A(\varphi,\varphi)$ and $B(\varphi,\varphi,\varphi)$ vanish,
i.e. \begin{eqnarray*}
\left.\frac{d}{dt}A\left(\varphi(t),\varphi(t)\right)\right|_{t=0} & = & 0\\
\left.\frac{d}{dt}B\left(\varphi(t),\varphi(t),\varphi(t)\right)\right|_{t=0} & = & 0\;,\end{eqnarray*}
 and for $B(\varphi,\varphi,\varphi)$ \begin{eqnarray*}
 &  & \left.\frac{d}{dt}B(\varphi(t),\varphi(t),\varphi(t))\right|_{t=0}\stackrel{!}{=}0\\
 & = & B([\psi,\varphi(0)],\varphi(0),\varphi(0))+B(\varphi(0),[\psi,\varphi(0)],\varphi(0))\\
 &  & +B(\varphi(0),\varphi(0),[\psi,\varphi(0)])\\
 & = & 3B([\psi,\varphi(0)],\varphi(0),\varphi(0))\\
 & = & 3B(\psi\cdot\varphi(0),\varphi(0),\varphi(0))-3B(\varphi(0)\cdot\psi,\varphi(0),\varphi(0))\\
 & = & 3(B(\psi,(\varphi(0))^{2},\varphi(0))-B(\psi,\varphi(0),(\varphi(0))^{2}))\qquad\mbox{ see (\ref{symmetric2})}\\
 & = & 0\end{eqnarray*}
 The last identity requires that $\varphi$ is a projector \begin{equation}
\varphi^{2}=\tau\varphi\,.\label{eq:projector}\end{equation}
Using that requirement we will show in the following that these projectors
are the basis elements of $\mathcal{T}$. The support of the curve
$\varphi(t)$ is a connected 3-manifold $\Sigma_{\varphi(t)}$. By
topological arguments about decompositions of 3-manifolds (see \cite{Mil:62,JacSha:79}),
we can always cut the support of a singular form into simple pieces.
We call the algebra elements of these simple pieces the basis of the
algebra. In the next section we will construct such a basis. The simple
pieces have intersecting supports which implies that all basis elements
fulfill relation (\ref{rel2}). By the linear independence, it is
enough to study the infinitesimal deformations for the basis elements
of the algebra. Thus we obtain the result: The invariance of the Bianchi
identity $B(\varphi,\varphi,\varphi)=0$ with respect to infinitesimal
deformations (\ref{deformation}) leads to the fact that the \emph{basis
elements} of the algebra $\mathcal{T}$ are \emph{projectors} (\ref{eq:projector}).
The case of deformation of $A(\varphi,\varphi)$ gives no new results
and it is not difficult to see that the expression $A(\varphi,\varphi)$
is an invariant of the algebra $\mathcal{T}$. 

The discussion above has shown that:

\begin{enumerate}
\item The product between two elements $\varphi,\:\psi\in\mathcal{T}$ is
divided into two cases: $\varphi\cdot\psi=\psi\cdot\varphi,\:(\varphi\cdot\psi)^{2}=\tau(\varphi\cdot\psi)$
and $\varphi\cdot\psi\cdot\varphi=\tau\varphi$, where all basis elements
are projection operators $\varphi^{2}=\tau\varphi$.
\item The first case corresponds to the disjoint union $C_{\varphi}\sqcup C_{\psi}$
of the corresponding curves $C_{\varphi},\, C_{\psi}$.
\item The second case describes the linking $C_{\varphi}\between C_{\psi}$
of the curves. All such elements $\varphi,\,\psi$ are non-commutative.
\end{enumerate}
The product of two elements $\varphi\cdot\psi$ is \emph{non-zero}
if the their supports \emph{intersect} $\Sigma_{\varphi}\cap\Sigma_{\psi}\neq\emptyset$.
If further the associated curves are \emph{linked} $C_{\varphi}\between C_{\psi}$
the elements are \emph{non-commuting} $\varphi\cdot\psi\neq\psi\cdot\varphi$.
In four dimensions the support of the differential form is a 3-dimensional
space and the intersection is a 2-dimensional.

\begin{figure}
\includegraphics[%
  scale=0.5]{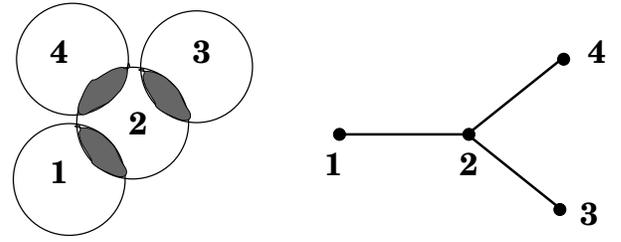}

\caption{Representation of a 3-manifold as tree\label{fig:1}}
\end{figure}

We can describe this in a compact form by a \emph{graph} with a vertex
for every 3-dimensional support and an edge between any two vertices
that correspond to intersecting supports. In the sequel we are going
to simplify the graph just described. To this end we need to employ
a certain fact about the structure of 3-manifolds. Every 3-manifold
is the boundary of an uniquely given 4-manifold. According to the
paper \cite{CuFrHsSt:97}, this 4-manifold must be contractible. But
this implies that the graph described above must be acyclic, since
otherwise the corresponding 3-manifold would be the boundary of a
non-contractible 4-manifold. An acyclic graph is a tree; thus, the
graph under consideration is a tree, which is finite due to the compactness
of the 3-manifold $\Sigma$. Our tree consists of two kinds of elements,
viz. branching vertices with more than two neighbours, and {}``regular''
vertices with one or two neighbours. A branching vertex corresponds
to the intersection of more than two 3-manifolds (see figure \ref{fig:1},
vertex 2 is the branching vertex). %
The intersection of two 3-manifolds embedded into a 4-manifold is
a 2-manifold. Hence, branching vertices in the graph corresponding
to a disjoint union of 2-manifolds. Two intersecting 2-manifolds generate
a cycle in the graph, which contradicts the simple-connectivity of
the 4-manifold. 

\begin{figure}
\includegraphics[%
  scale=0.5]{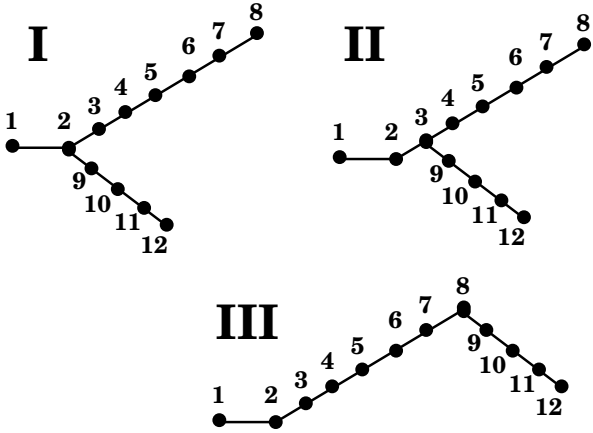}

\caption{Fig. I is the starting tree. Fig. II shows the modification of the
tree after the resolving of the branching point 2 which is moved to
the point 3. Fig. III is the result of the repetition of the procedure
resulting in a line.\label{fig:2}}
\end{figure}

Next we will describe a procedure which converts the branching vertices
to regular vertices by changing the 3-manifolds. By diffeomorphisms
we can move the 2-manifolds. When doing so it can happen that we produce
intersections of 2-manifolds. According to a result of \noun{Casson}
\cite{Kir:89} it is possible to deform the 2-manifolds to remove
these intersections at the price of resulting self-intersections of
the deformed 2-manifolds. By this procedure we can rebuild the tree
(see figure \ref{fig:2}), %
and one branching vertex is converted to a regular one. By this procedure,
the structure of the 3-manifold will be modified. We will discuss
this fact in a later paper. A repetition of this procedure amounts
to shifting a subtree along some path down to some leaf of the tree.
By doing so, the vertex where the subtree was rooted originally is
turned to a regular (i.e. non-branching) vertex. Finally we can deform
the graph structure to a line. But that is an \emph{order --} the
order of the basis elements of the algebra $\mathcal{T}$. Thus, there
is a finite number of generators of $\mathcal{T}$ which are ordered
along a line. This is equivalent to a labeling of the generators by
integer numbers. We denote the generators by $e_{k}$ with $k\in\mathbb{N}$.

Putting all facts together we can construct the algebra of transitions
of the differential structure: To every support of a singular 1-form
we associate a generator $e_{i}$ of $\mathcal{T}$ which is a projection
operator up to a constant factor $\tau$, i.e. $e_{i}^{2}=\tau e_{i}$.
By the argumentation above, we can choose the order of the generators
in such a manner that the curves $C(e_{i})$ and $C(e_{j})$ of $e_{i}$
and $e_{j}$, respectively, are linked only for $|i-j|=1$. By definition
we express the relation for non-linked curves by $e_{i}e_{j}=e_{j}e_{i}$
for $|i-j|>\nobreak1$. Relation (\ref{rel2}) can be rewritten as
$e_{i}e_{i+1}e_{i}=\tau e_{i}$ or $e_{i+1}e_{i}e_{i+1}=\tau e_{i+1}$.
What remains to finish the proof of the theorem is the construction
of a $*$-operation and the establishment of the fact that the number
field is that of the complex numbers. 

The construction of the $*$-operation in $\mathcal{T}$ has been
done above by changing the orientation of the curve $C_{\varphi}$
for an element $\varphi\in\mathcal{T}$. Then the main relations like
$(\varphi^{*})^{*}=\varphi$ and $(\varphi\cdot\psi)^{*}=\psi^{*}\cdot\varphi^{*}$can
be justified. In \cite{Akb:99}, Akbulut constructed the support of
a singular 1-form for simple pieces corresponding to the base elements
$e_{i}$ of the algebra $\mathcal{T}$. Then he showed that the mirror
$C(e_{i}^{*})$ of the curve $C(e_{i})$ leads to a diffeomorphic
support of the singular 1-form $e_{i}^{*}$, i.e. the base elements
are self-adjoint $e_{i}^{*}=e_{i}$. 

\textcolor{black}{To show that the number field of $\mathcal{T}$
is that of the complex numbers we associate a singular connection
1-form $\varphi=f_{*}^{-1}df_{*}$ to a smooth, singular map $f\colon M\to N$
between two manifolds $M,\: N$ with different differential structures.
The curvature $\Omega$ is given by $\Omega=d(f_{*}^{-1}df_{*})+f_{*}^{-1}df_{*}\wedge f_{*}^{-1}df_{*}$
with support a surface $supp(\Omega)=S$ embedded into the 3-manifold
$\Sigma$ which is the singular set of $f$. To construct the surface,
we consider the Euler class $e(\Omega)$ associated to the curvature
2-form via the relation $e(\Omega)\wedge e(\Omega)=p_{1}(\Omega)$
with the first Pontrjagin class $p_{1}$. Then the surface $S$ is
defined by the relation \[
\int\limits _{S}e(\Omega)=1\]
 which defines a bilinear map $H_{2}(M,\mathbb{Z})\times H^{2}(M,\mathbb{R})\rightarrow\mathbb{R}$
between the homology class $[S]\in H_{2}(M,\mathbb{Z})$ of the surface
and the cohomology class $[\Omega]\in H^{2}(M,\mathbb{R})$ of the
curvature. The classification theory of vector bundles (see \cite{Hus:94}
for instance) states that every cohomology class in $H^{2}(M,\mathbb{R})$
is given by the curvature of a complex line bundle $L$ over the 3-manifold
$\Sigma$. Furthermore, let $s\colon\Sigma\rightarrow L$ be a section
of the complex line bundle, then the zero set ${x\in\Sigma\,|\: s(x)=(x,0)}$
is the homology class $[S]\in H_{2}(M,\mathbb{Z})$. Thus, the class
$[tr(\Omega)]$ is the curvature 2-form of a complex line bundle,
i.e. a differential 2-form with values in the Lie algebra $i\mathbb{R}$
of the $U(1)$ group. Let $\varphi$ be the singular 1-form which
generate via $\varphi L\varphi$ the curvature $\Omega$. For simplicity
we assume that the form $\varphi$ is given by $\varphi=c_{i}e_{i}$
with the base element $e_{i}$. Then, we obtain the coefficient of
the curvature to be \[
\Omega=\varphi L\varphi=c_{i}^{2}e_{i}Le_{i}\:.\]
 Any curvature 2-form $\Omega$ can be decomposed as a 2-form by $\Omega=\Omega_{\mu\nu}dx^{\mu}\wedge dy^{\nu}$
where $\Omega_{\mu\nu}$ is an anti-symmetric matrix with values in
$i\mathbb{R}$. A direct identification of the coefficient $c_{i}$
is given by the integral\[
\int\limits _{S}tr(\Omega)=Tr(tr(\Omega))=c_{i}^{2}\, Tr(e_{i}\, Le_{i})\:,\]
 and we obtain\[
c_{i}=\sqrt{Tr(tr(\Omega))}\in\mathbb{C}\;.\]
 Thus, the number field of the algebra is that of the complex numbers.
That completes the proof of the theorem.}

It remains to show that $\mathcal{T}$ can be defined as an algebra
of linear operators over some Hilbert space. This can be done by the
GNS construction (see \cite{Mur:90}). First we remark that the algebra
itself is also a vector space. Given two elements $\varphi,\psi\in\mathcal{T}$
then $Tr(\psi^{*}\varphi)=\left\langle \psi,\varphi\right\rangle $
is the \emph{scalar product}. The \emph{self-adjoint} elements $e_{k}$
are projection operators and are the \emph{basis} elements of the
Hilbert space. Finally it follows: The completion of the algebra $(\mathcal{T},Tr)$
is a \emph{complex Hilbert space.} By fixing the parameter $\tau$
of the algebra $\mathcal{T}$ to be $\tau=1/2$, the completion of
$\mathcal{T}$ corresponds to the \emph{Fock space of fermions} in
quantum field theory (see \cite{PlyRob:94}, chapter 2). That is a
remarkable result: \emph{The self-adjoint projectors} $e_{k}$ \emph{generate
the creation and annihilation operators of the fermions.} That means,
for $\tau=1/2$ the algebra $\mathcal{T}$ is the standard Clifford
algebra of anti-commutative operators. For the case $\tau\not=1/2$,
$\mathcal{T}$ extends the standard quantum field algebra to a Temperley
Lieb algebra.

The algebra $\mathcal{T}$ was introduced by \textsc{Temperley} and
\noun{Lieb} in the context of statistical mechanics and by \noun{Jones}
for the description of the so-called hyperfinite factor $II_{1}$
$C^{*}$-Algebra. Later this algebra was used by \noun{Kauffman}
and \noun{Lickorish} to define invariants of a three-manifold identical
to the so-called quantum invariants, and we will discuss this interesting
relation in the last section.

\section{Space-Time realization of the field operator algebra}

\label{sec:4}Using the results of the previous section it is possible
to construct a concrete realization of the algebra (\ref{Jones-algebra})
of transitions of the differential structure of space-time. Let us
start with the question: Which operations change the differential
structure of space-time but do not alter the topology? Up to the present
two such {}``surgeries'' are known: the logarithmic transform and
its generalization -- the knot surgery. In this section we will show
that all generators $e_{2i-1}$ for $i>0$ are represented by the
logarithmic transform, whereas all possible generators $e_{i}$ corresponding
to the knot surgery. 

In a short overview both operations can be described easily. As a
pre-requisite we need a topologically complicate four-manifold. The
\emph{logarithmic transform} is a procedure removing a torus with
neighborhood and sewing in a twisted version of the torus, i.e. we
cut the torus, twist one end $p$ times and glue both ends together.
As proved by Gompf \cite{Gom:91}, the new four-manifold has a different
differential structure for $p>0$. Later Fintushel and Stern \cite{FiSt:96}
generalize this operation to the \emph{knot surgery}: Instead of using
only twists of the end, the end can be knotted according to a knot.
Fintushel and Stern show that for different knots, the resulting four-manifolds
may have different differential structures. Furthermore it was conjectured
that any two different knots $K_{1},\: K_{2}$ always lead to different
differential structures. Later, this conjecture was corrected by Akbulut
\cite{Akb:99}: The knot $K_{1}$ and its mirror $\overline{K}_{1}$
have to be different to the knot $K_{2}$, then the corresponding
differential structures are also different.

Now we will describe the technical details of both constructions.
A good point to begin with is to consider the simplest transition
of the differential structure: the logarithmic transform representing
the commuting subset of generators of $\mathcal{T}$. Given a sufficiently
non-trivial simply-connected four-manifold %
\footnote{The four-manifold $M$ must satisfied the condition $b_{2}^{+}(M)>0$.%
}. A \emph{logarithmic transform} is a transition of the four-manifold
$M$ by cutting out a neighborhood $N(T^{2})$ of an c-embedded %
\footnote{We call a torus $T$ \emph{c-embedded} in $M$, if there is a neighborhood
$N(T)$ in $M$ and a diffeomorphism $\phi:N(T)\rightarrow U$ where
$U$ is a neighborhood of a cusp fiber in an elliptic surface and
$\phi(T)$ is a smooth elliptic fiber in $U$.%
} torus $T^{2}=S^{1}\times S^{1}$ with no self-intersections and sewing
in a twisted torus. In the following we will construct the algebra
$\mathcal{A}$ of transitions of the differential structure of $M$
induced by logarithmic transforms.

Consider a neighborhood $N(T)=D^{2}\times T$ of an c-embedded torus
$T$ in $M$ with $\pi_{1}(M\setminus T)=0$. We remove $N(T)$ from
$M$ to get $M\setminus N(T)$ with boundary $\partial(M\setminus N(T))=T^{3}$.
We glue it back by an orientation-reversing diffeomorphism $g\colon\partial N(T)=S^{1}\times T\rightarrow\partial(M\setminus N(T))$.
For every point $x\in T$ we have the map $g(.,x)=\tilde{g}\colon S^{1}\rightarrow S^{1}$
given by $z\mapsto z^{p}$ for an integer $p>0$, also called multiplicity.
A main theorem of \textsc{Gompf} \cite{Gom:91} states that the altered
manifold \[
{M}_{p}=(M\setminus N(T))\cup_{g}N(T)\]
 is homeomorphic to $M$ but \emph{not} diffeomorphic for all $p>\nobreak1$.

Both surgery steps can be performed simultaneously on $k$ c-embedded
tori: Taking $k$ c-embedded tori $T_{1},...,T_{k}$, $k$ nonnegative
integers $p_{1},...,p_{k}$, and choosing diffeomorphisms $g_{i}\colon\partial N(T_{i})=S^{1}\times T_{i}\rightarrow\partial(M\setminus N(T_{i}))$
with multiplicity $p_{i}$. The corresponding four-manifold $M_{p_{1}\ldots p_{k}}$
is homeomorphic to $M$ if the multiplicities are coprime, i.e. there
is no common divisor, but not diffeomorphic. Now we associate $k$
generators of $\mathcal{T}$ to each of these transformed neighborhoods
$N(F_{i})_{p_{i}}$. For a subset of $\mathcal{T}$ with $s=2i-1\,,\,\, i=1,\ldots,k$
the generators $e_{s}$ commute and we can form the \emph{superposition}\begin{equation}
\varphi=\sum\limits _{i=1}^{k}a_{i}(p_{i})e_{2i-1}\label{log-Hilbert}\end{equation}
 with coefficients $a_{i}(p_{i})$ depending only on the multiplicities.
The corresponding curvatures are also superpositions \[
\Omega=\sum\limits _{i=1}^{k}(a_{i}(p_{i}))^{2}[e_{2i-1}Le_{2i-1}]\]
 representing the disjointness of the collection of neighborhoods
$N(F_{i})_{p_{i}}$.

The general case for all generators can be treated by a construction
of \noun{Fintushel} and \noun{Stern} generalizing the logarithmic
transform. Let $X$ be a simply connected smooth four-manifold which
contains a smoothly embedded torus $T$ with self-intersection $0$.
Given a knot $K$ in $S^{3}$, we replace a tubular neighborhood of
$T$ with $S^{1}\times(S^{3}\setminus K)$ to obtain the \textit{knot
surgery manifold} $X_{K}$. In other words, we replace the torus by
knotted torus according to the knot $K$. Of course one can generalize
this procedure by using $n$ embedded tori with self-intersection
$0$ and a link $L$ with $n$ components (for the details see \cite{FiSt:96}).
The main point is that a link with $n$ components can be interpreted
as a composition of two overlapping neighborhoods $N(F_{1})_{p_{1}}$
and $N(F_{2})_{p_{2}}$ described by logarithmic transforms. The multiplicities
now correspond to the framings of the link components. The generators
do \emph{not} commute and we obtain the general superposition \begin{equation}
\varphi=\sum\limits _{i=1}^{n}a_{i}(L_{i})e_{i}\label{link-Hilbert}\end{equation}
 with coefficients depending only on the link components $L_{i}$.

By means of the knot surgery the \emph{$*$ operation} can be defined
by the construction of the mirror link $-L$ related to link $L$
inducing a $\varphi^{*}$ for each $\varphi$. In \cite{Akb:99} \noun{Akbulut}
shows by an explicit construction of the handle body structure of
$X_{L}$ using the handle body structure of $X$ that $X_{L}$ and
$X_{-L}$ must be diffeomorphic. Thus the generators $e_{i}$ have
to be \emph{self-adjoint} $e_{i}^{*}=e_{i}$. This emphasizes the
construction of the $*$-operation in the $C^{*}$-algebra $\mathcal{T}$
of the transitions of the differential structure.\\
\\
The fact that quantum mechanics requires a complex \emph{}Hilbert
space has been a deep mystery since the very early days of quantum
theory. No satisfying explanation could be found, and most text books
discount this question as an inconsequential detail of the technical
calculus. But the truth is that without the complex structure quantum
mechanics would miss inherent properties which are responsible for
the term {}``quantum'' in its name, e.g. the interference terms
of a superposition or the oscillating solutions of the Schrödinger
equation. In Section \ref{sec:3} we constructed the coefficient field
of the $C^{*}$-algebra $\mathcal{T}$ of transitions of differential
structures. It was shown that the coefficient field are the complex
numbers and thus the space of the differential structures is by necessity
a complex Hilbert space. In the following we will give an explicit
construction of the amplitudes and phases of the quantum states for
the logarithmic transform.

At first we have to consider the transition of the differential structure
again. Let us start with the case of a logarithmic transform, i.e.
we remove the neighborhood $N(T)=T\times D^{2}$ of an embedded torus
$T$ in the 4-manifold $M$ and glue it back by using a diffeomorphism
$g\colon\partial N(T)\to\partial(M\backslash N(T))$ to generate the
non-diffeomorphic manifold $M_{g}$. Thus, we have to study this map
$g$ for the calculation of the change of the connection. We remark
the decomposition of $g$ into $(id_{T},\:\tilde{g})$ with $\tilde{g}\colon S^{1}\to S^{1}$
defined by $z\mapsto z^{p}$ for $z\in S^{1}\subset\mathbb{C}$ and
we call $p$ the multiplicity of $g$. The details of the following
calculation will be published in the paper \cite{AssBra:05}. Let
$e,\, f$ be the reference frames for the tangent bundles $TM,\: TM_{g}$,
respectively. It is easy to see that $g$ can be extended to a smooth
map $f\colon M\to M_{g}$ with the differential $df\colon TM\to TM_{g}$.
The differential is a mapping of the reference frames given by $df(e)=a\cdot f$
with the function $a\colon M\to G$ where $G$ is the structure group
of the tangent bundle $TM$. The map $a$ is the identity outside
of $N(T)$ and the restriction to $\tilde{g}$ is given by $a(z)|_{\tilde{g}}=z^{1-p}/p$
where $z\in D^{2}\subset\mathbb{C}$. Let $\omega_{e},\:\omega_{f}$
be the connections on $TM,\, TM_{g}$, respectively. Finally we obtain\[
\omega_{e}=df^{-1}\omega_{f}df+a^{-1}da\]
 with the form $a^{-1}da$ singular along the torus at $z=0$. Only
the map $\tilde{g}$ is singular leading to the form\[
a^{-1}da=-\frac{p-1}{p}\,\frac{dz}{z}\]
 defining the curvature\begin{equation}
d(a^{-1}da)=-\frac{p-1}{p}\:\delta(z)\, dz\wedge d\overline{z}\label{log-trafo-curvature}\end{equation}
 with Diracs delta function $\delta(z)$. The singular form $\varphi(z)=\delta(z)dz\wedge d\overline{z}$
is purely imaginary, i.e. $\overline{\varphi(z)}=-\varphi(z)$ and
can be represented by $\varphi=i\:\Im\varphi$. The integral over
that form is equal to one and the coefficient of the curvature change
is given by \[
{i}\,\frac{p-1}{p}\,.\]
 Two questions remain: Where does the imaginary unit really comes
from and what is the 3-manifold $\Sigma$? As stated above, we have
a singularity at the origin $z=0$ of $D^{2}$ in the neighborhood
$N(T)=T\times D^{2}$ of the torus $T$. This singularity is described
by the complex function $z^{p}$. By using complicate methods from
singularity theory, one can construct a knot in $S^{3}$ which modifies
$S^{3}$ by surgery to a homology 3-sphere $\Sigma$, i.e a 3-manifold
with the same homology as $S^{3}$ but different fundamental group.
A construction of the 3-manifold $\Sigma$ for the case of a logarithmic
transform can be found in \cite{Gom:91}. Then consider a $SO(3)$
bundle over $\partial N(T)=T^{3}$. By using fiber bundle theory,
every $SO(3)$ bundle is completely classified by the Euler class
$e(M)$, i.e. a 2-form which is closed $de(M)=0$ but not exact $e(M)\not=d\varphi$
(for all 1-forms $\varphi$). But that Euler class is induced by a
non-trivial $SO(2)$ bundle over $T^{3}$ because the connection of
the $SO(3)$ bundle has vanishing curvature. Thus the non-trivial
curvature (\ref{log-trafo-curvature}) is the curvature of the $SO(2)$
bundle. But $SO(2)$ is isomorphic to $U(1)$ and the Lie algebra
of $U(1)$ are the pure imaginary numbers $i\mathbb{R}$. That is
where the imaginary unit comes into play: As part of the only non-trivial
bundle in dimension~3, the $U(1)$ bundle over the boundary $\partial N(T)=T^{3}$.

By using that result, let us look at the structure group of the tangent
bundle $TM$ of a four-manifold $M$. In the case of a Riemannian
manifold the structure group is $SO(4)$ whereas in the Lorentzian
case we get the Lorentz group $SO(3,1)$. Both groups have Lie algebras
which are isomorphic to the Lie algebras of the corresponding spin
groups, i.e. $SU(2)\times SU(2)$ for the Riemannian and $SL(2,{\mathbb{C}})$
for the Lorentzian case. The Lie algebra consists of all anti-hermitian
operators in ${\mathbb{C}}^{2}$, i.e. operators of the form $iH$
where $H$ is a hermitean operator in ${\mathbb{C}}^{2}$. The exponential
relates the Lie algebra to the Lie group. Thus, the coefficients in
(\ref{link-Hilbert}) are elements of the Lie group and we obtain\[
a_{i}(L_{i})=A_{i}\exp\left(2\pi i\frac{p_{i}-1}{p_{i}}\right)=A_{i}\exp\left(-2\pi i\frac{1}{p_{i}}\right)\,,\]
where $p_{i}>1$ is the \emph{framing number} of the link component
$L_{i}$ and $A_{i}$ is a normalization constant. We call $A_{i}$
the \emph{amplitude} and $2\pi/p_{i}$ the \emph{phase}. The star
operation changes the knot to its mirror which changes the framing
from $p_{i}$ to $-p_{i}$. Thus the $a_{i}$ are complex conjugated
by using the star operation $a_{i}^{*}$. The normalization of $A_{i}$
is given by \begin{eqnarray*}
1=\sum\limits _{i}A_{i}^{2}Tr(e_{i}^{2})\end{eqnarray*}
 There are two reasons for this normalization: At first the trace
of an element of $\mathcal{T}$ must lie in the interval $[0,1]$
by definition and without loss of generality we normalize the trace
to $1$. Secondly, if we compare two different differential structures
described by the same basis elements in $\mathcal{T}$ then we need
a normalization to compare the differential structures by comparing
the coefficients.

\section{The connection to Loop Quantum Gravity and Summary}

\label{sec:6}Before we discuss the consequences of our approach we
will collect the results of the paper:

\begin{enumerate}
\item The set of differential structures on a space-time is the set of non-diffeomorphic
reference frames. Matter is the transition between reference frames
that belong to different differential structures of the space-time.
\item The transition of the differential structure between two topologically
equivalent 4-manifolds $M,N$ can be described by a smooth map $f\colon M\to N$
having singularities along a 3-manifold $\Sigma$.
\item We have constructed a $C^{*}$ algebra $\mathcal{T}$, the Temperley-Lieb
algebra, representing the set of operators of transitions of the differential
structure of space-time. Furthermore we are able to implement the
structure of a complex Hilbert space on the set of differential structures
by using the GNS construction.  
\item The transition of the differential structure is a modification of
a contractable 4-dimensional subset of the 4-manifold. This subset
has a boundary, the 3-manifold $\Sigma$. We associated to every transition
an element of the algebra $\mathcal{T}$. 
\item We have constructed a complex structure on the algebra $\mathcal{T}$
by using the structure group of the tangent bundle of $\Sigma$.
\end{enumerate}
At the end of section \ref{Theorem1} we remarked that in a series
of papers, \noun{Lickorish} and \noun{Kaufman} used the Temperley-Lieb
algebra to construct knot invariants in 3-manifolds. If we go further
into that direction then we will expect that knots and links are also
important in our theory. A first step in that direction is the space-time
realization of the generators of the algebra $\mathcal{T}$ by {}``surgeries
along knots and links'' introduced by \noun{Fintushel} and \noun{Stern}.
But there is a very popular approach to quantum gravity which uses
extensively knots and links, \emph{Loop quantum gravity}. 

Loop quantum gravity is based on the formulation of classical general
relativity, which goes under the name of {}``new variables'', or
\noun{{}``Ashtekar} variables\noun{''} (see \cite{Ash:86}).
Soon after the introduction of the new variables, it was realized
that the reformulated Wheeler-DeWitt equation admits a simple class
of exact solutions: the trace of the holonomy of the Ashtekar connection
around smooth, non-self-intersecting loops. Then \noun{Rovelli}
and \noun{Smolin} \cite{RoSm:88} choose these Wilson loops as the
new basis in the Hilbert space of the theory. 

Loop quantum gravity is the formulation of canonical general relativity
by a $SU(2)$ connection $A$ on a principal bundle over a 3-manifold
and a tetrad field (also called the soldering form) of density weight
one also defined on a 3-manifold. This approach assumes the existence
of a global Lorentzian structure on the 4-manifold by introducing
a global foliation $\Sigma\times\mathbb{R}$ of the 4-manifold. Then
a graph $\Gamma$, i.e. a finite collection of smooth oriented 1-dimensional
submanifolds also called links, in $\Sigma$ is chosen with $L(\Gamma)$
components overlapping only at the endpoints also called nodes. Given
a $SU(2)$ connection $A$, the holonomy $U_{l}(A)\in SU(2)$ of the
connection $A$ along the link $l$ is defined to be\begin{equation}
U_{l}(A)=\mathcal{P}\:\exp\left(\int_{l}\, A\right)\label{loop-variable}\end{equation}
 where $\mathcal{P}$ denotes path ordering. Then a graph $\Gamma$
induces a map $A\mapsto\prod_{l\in L(\Gamma)}(U_{l}(A))=p_{\Gamma}(A)$
which defines via a function $h\colon[SU(2)]^{\times L(\Gamma)}\to\mathbb{C}$
the so-called cylindrical function by\[
\Psi_{\Gamma,h}(A)=h\left(U_{l_{1}},\ldots,U_{l_{L(\Gamma)}}\right)\:.\]
 Quantum states in Loop quantum gravity are limit sequences of cylindrical
functions with respect to the $L^{2}$ norm. In \cite{FaiRov:04},
\noun{Fairbairn} and \noun{Rovelli} show that the Hilbert space
of diffeomorphism classes of such graphs is non-separable. Furthermore
both authors construct a new, separable Hilbert space by extending
the diffeomorphisms to smooth maps with a finite number of singularities.
That is the first hint that quantum gravity could incorporate differential
structures. But there is more. 

In the introduction we have constructed a singular connection representing
the transition of the differential structure. Mathematically such
singular connections are distribution-valued differential forms \cite{Ass:96}.
The space of connections including such singular connections is the
4-dimensional analog of the space $\overline{\mathcal{A}}$ in Loop
quantum gravity and the use of the trace $Tr(D)$ for the connection
corresponds to the gauge equivalence classes $\overline{\mathcal{A}/\mathcal{G}}$.
If that correspondence is more than an accident we have to interpret
the holonomies $U_{l}(A)$ and the graph $\Gamma$ in a 4-dimensional
context. Here we will give a sketch with such an interpretation, where
the details will appear in a follow-up paper.

Given a 3-manifold $\Sigma$ and a $SL(2,\mathbb{C})$ bundle over
the 4-manifold $M$. The 3-manifold is the singular set of a map,
which changes the differential structure on $M$. Let $\tilde{A}$
be a $SL(2,\mathbb{C})$ connection on a neighborhood $N(\Sigma)$
of the embedded 3-manifold which restricts to the $SU(2)$ connection
$A$ on $\Sigma$. By a bundle theoretic reduction, this restriction
is unique. As usual, the graph in $\Sigma$ is a subset of the triangulation
of $\Sigma$. All possible graphs in $\Sigma$ can be interpreted
as the set $P\Sigma$ of all paths in $\Sigma$. Now we define a map
$\pi\colon P\Sigma\to\Sigma$ which assigns to each path its end point.
That defines a fibration of the space $P\Sigma$ into the base $\Sigma$
and fiber $\Omega\Sigma$, the space of closed loops in $\Sigma$.
Thus the path space $P\Sigma$ is locally generated by the loop space
$\Omega\Sigma$ and in the following we will mainly work in that space. 

The topology of the loop space is non-trivial, and we are interested
in the connected components of $\Omega\Sigma$ which is the fundamental
group $\pi_{1}(\Sigma)$, i.e. the group of loops up to continuous
homotopy. Why are we interested in that group? Let us consider the
holonomy $U_{\gamma}(\tilde{A})$ of a $SL(2,\mathbb{C})$ connection
over $\Sigma$ along a loop $\gamma$. It is easy to show that $U_{\gamma}(\tilde{A})$
only depends on the homotopy class of the loop $\gamma$, i.e. given
two homotopic loops $\gamma,\psi$, then $U_{\gamma}(\tilde{A})=U_{\psi}(\tilde{A})$.
Thus the holonomy defines a representation $U\colon\pi_{1}(\Sigma)\to SL(2,\mathbb{C})$
of the fundamental group into $SL(2,\mathbb{C})$. What does the representation
mean? The group $SL(2,\mathbb{C})$ is the spin group on the Minkowski
space, and every representation in that group is equivalent to a representation
into the Lorentz group $SO(3,1)$ and vice versa. Thus the holonomy
defines also a representation $U\colon\pi_{1}(\Sigma)\to SO(3,1)$. 

The Lorentz group $SO(3,1)$ is the isometry group of the 3-dimensional
hyperbolic space $\mathbb{H}_{\infty}^{3}$ and the representation
$U$ defines a hyperbolic geometry a la \noun{Thurston} \cite{Thu:97}
with finite volume on the 3-manifold $\Sigma$. Two representations
$U_{1},U_{2}$ which are not related via conjugation to each other
define different hyperbolic structures. In the next paper we will
show that \emph{the hyperbolic structures on a 3-manifold are determined
by the differential structure of the 4-manifold in which the 3-manifold
embeds}. Thus, the holonomies over the loops in the 3-manifold generate
the hyperbolic structure on the 3-manifold and are related to the
differential structures on the 4-manifold. An evolution of the 3-manifold
$\Sigma$ which changes the hyperbolic structure results in a conical
singularity as firstly observed in the spin foam models.

Another open point in the Loop quantum gravity is the foliation of
space-time breaking the general covariance of loop theory. With the
differential structure approach we are able to determine such a splitting
in space and time naturally -- at least locally. In the neighborhood
of the 3-dimensional singular support $\Sigma$ we have a \emph{canonical
splitting} of the four-manifold into a 3-dimensional and a one-dimensional
space. Given two singular supports $\Sigma_{1},\:\Sigma_{2}$ which
do not intersect then two splittings are mostly \emph{not} identical.
That would be a possible explanation of the appearance of singular
knots and conical singularities in loop quantum gravity. From the
understanding of these points, it should be possible to obtain a full
solution of the Hamilton constraint by using the methods above more
extensively. In the next paper we will go further and describe something
like the dynamics of differential structures and its localization
to 3-manifolds.

\section*{\textcolor{black}{Appendix}}

\textcolor{black}{In the following we will explain the definition
of a differential $df$ of a singular map $f$ by using the work of
Harvey and Lawson \cite{HarLaw:93}. At first we describe the general
situation.}

\textcolor{black}{Let $F$ and $E$ be two vector bundles over $M$.
Furthermore, let $\alpha:E\rightarrow F$ be a bundle map admitting
singularities, i.e. a subset $\Sigma\subset M$ where the (local)
map $\alpha_{x}:E_{x}\rightarrow F_{x}$ is not injective. Let $D_{E}$
and $D_{F}$ be the connections on $E$ and $F$, respectively. The
main problem is now that the inverse of the map $\alpha$ not always
exists. Instead we have to define a map $\beta:F\rightarrow E$. At
first we assume such a map exists. Then we define the push forward
connection $\vec{D}^{\alpha}$ by \[
\vec{D}^{\alpha}=\alpha\circ D_{E}\circ\beta+D_{F}\circ(1-\alpha\beta)\,.\]
 On the two bundles we introduce metrics allowing to define the adjoint
$\alpha^{*}$ of $\alpha$ via the scalar product. Then outside of
the singular set $\Sigma$ we define \[
\beta=(\alpha^{*}\alpha)^{-1}\alpha^{*}\,.\]
 In general this procedure breaks down on the singular set $\Sigma$,
since $\beta$ becomes singular on $\Sigma$. To change this, we choose
an approximation mode, i.e. a fixed smooth function $\chi:[0,\infty]\rightarrow[0,1]$
with $\chi'\leq0,\chi(0)=0$ and $\chi(\infty)=1$. We define a smooth
approximation $\beta_{s}$ to $\beta$ by \[
\beta_{s}=\chi\left(\frac{\alpha^{*}\alpha}{s}\right)\beta\qquad\mbox{for $s>0$}\,.\]
 This family $\beta_{s}$ of maps defines a family of connections
$\vec{D}_{s}^{\alpha}$ on $F$. As $s\rightarrow0$ the map $\beta_{s}$
converge to $\beta$ uniformily on compacta in $M-\Sigma$. The family
$\beta_{s}$ of maps converges for $s\rightarrow\infty$ pointwise
to a connection on $F$ for all points in $M$ (see \cite{HarLaw:93}
for the proofs).}

\textcolor{black}{This general situation can be used to define the
connection change by setting $E=TM$, $F=TN$ and $\alpha=df$. Then
the substitute for the inverse $df^{-1}$ is defined by the limit
$s\rightarrow\infty$ of $\beta_{s}$ constructed above. }

\begin{acknowledgments}
\textcolor{black}{We thank Carl H. Brans for valuable discussion about
the relevance of differential structures in physics. Special thanks
to Andreas Schramm for important corrections. We thank the members
of the physicsforums (http://www.physicsforums.com) for criticism
and helpful remarks.}
\end{acknowledgments}

\end{document}